\documentclass{PoS}

\title{Chiral symmetry of graphene and strong coupling lattice gauge theory}

\ShortTitle{Chiral symmetry of graphene and strong coupling lattice gauge theory}

\author{\speaker{Yasufumi Araki}$^{(a)}$ and Tetsuo Hatsuda$^{(a,b)}$ \\
   $ ^{(a)}$ Department of Physics, The University of Tokyo, Tokyo 113-0033, Japan\\
   $ ^{(b)}$ Institute for the Physics and Mathematics of the Universe (IPMU),
    The University of Tokyo, Chiba 277-8568, Japan\\
  E-mail: \email{araki@nt.phys.s.u-tokyo.ac.jp}, \email{hatsuda@phys.s.u-tokyo.ac.jp}}

\abstract{
 We model the electrons on a monolayer graphene
  in terms of the   compact and non-compact  U(1) lattice gauge theories. 
  The system is analyzed by the  strong coupling expansion and 
  is shown to be an insulator due to dynamical gap formation
   in/around the strong coupling limit.
 This is similar to the spontaneous chiral symmetry breaking
 in strong coupling gauge theories.
 The results from the compact and  non-compact formulations
 are compared up to the next-to-leading order of the strong coupling expansion.
 Excitonic modes and their dispersion relations in the insulating phase
 are also investigated: it is found that there arises  
 a pseudo-Nambu--Goldstone mode obeying the Gell-Mann--Oakes--Renner type
 formula.}

\FullConference{The XXVIII International Symposium on Lattice Field Theory, Lattice2010\\
		June 14-19, 2010\\
		Villasimius, Italy}

\begin{document}

\section{Introduction}
After its first experimental discovery in 2004 \cite{Novoselov_2004},
graphene (monoatomic layer material of carbon atoms)
has widely attracted theoretical and experimental 
attention \cite{castro_neto_2009}.
Due to its hexagonal lattice structure,
 charge carriers on a graphene reveal a linear dispersion relation
around two ``Dirac points'' in momentum space \cite{wallace_1947},
 so that quasiparticles at low energies 
can be described as two species of massless Dirac fermions
 in (2+1)-dimensions \cite{Semenoff:1984dq}.
The symmetry between two triangular sublattices 
of graphene is referred to as the ``chiral symmetry''.

In the low-energy effective  theory of graphene,
 the effective Coulomb interaction between charge carriers
 is enhanced by 300 times due to the  small Fermi velocity $v_{_F}$.
Such a strong Coulomb interaction may turn a suspended graphene in the vacuum
from semimetal to insulator by the formation of a finite spectral gap of quasiparticles \cite{CN09}.
This mechanism is similar to the spontaneous chiral symmetry breaking
and associated fermion mass generation in quantum chromodynamics (QCD).
Various attempts have been made so far
to study the gap formation in monolayer graphene by 
using  the Schwinger--Dyson equation \cite{khveshchenko_2001,gorbar_2002,gorbar_2009},
the $1/N$ expansion \cite{Herbut_2006,son_2007,son_2008},
the exact renormalization group \cite{Giuliani_2010},
and the lattice Monte Carlo simulations \cite{hands_2008,drut_2009,giedt_2009,drut_2010}.
 These works are mainly focused on the critical region of semimetal-insulator
  transition or on the behavior for large number of flavors.
 
In this work, we rather focus on the strong coupling region 
 of the system and study  the low-energy effective theory
 discretized on a square lattice \cite{Araki_Hatsuda_2010,Araki_2010}.
 By using the strong coupling expansion of the 
  compact and non-compact formulations
  of the gauge field,
 we study the gap formation due to the spontaneous ``chiral symmetry'' breaking
  as well as the dispersion relations for collective excitations. 
 Typical energy scale of the emergent excitations are also 
 estimated with the use of the intrinsic length scale  of the original honeycomb lattice.

\section{Low-energy effective theory of graphene}
\subsection{Effective action in the continuum limit}
With the annihilation operators of the electrons on the 
two triangular sublattices of graphene ($a_\sigma$ and $b_\sigma$) near the two Dirac points $\mathbf{K}_\pm$,
we can construct a 4-component spinor in the momentum space,
$\psi_\sigma(\mathbf{p}) \equiv \left(a_\sigma(\mathbf{K}_+ + \mathbf{p}),b_\sigma(\mathbf{K}_+ + \mathbf{p}),b_\sigma(\mathbf{K}_- + \mathbf{p}),a_\sigma(\mathbf{K}_- + \mathbf{p})\right)^T$.
Here $\sigma=\uparrow,\downarrow$ denotes the original spin of the electrons.
The Euclidean effective action for graphene is then written as \cite{gorbar_2002,son_2007}
\begin{equation}
S_E = \sum_{\sigma}
 \int dx^{(3)} \ \bar{\psi}_\sigma \left( D[A] +m \right) \psi_\sigma + \frac{1}{2g^2} \sum_{j=1,2,3}  \int dx^{(4)} (\partial_j A_4)^2 ,
 \label{eq:effaction}
\end{equation}
where the Dirac operator reads $D[A]= \gamma_4(\partial_4+iA_4) + v_{_F} \sum_{i=1,2} \gamma_i \partial_i$.
This is  analogous to the action in ``reduced QED'' \cite{gorbar_2001},
in which the fermion $\psi$ in (2+1)-dimensions
 is interacting with the U(1) gauge field $A$
in  (3+1)-dimensions.
The Hermitian $\gamma$ matrices obey the anticommutation relation $\{\gamma_\mu,\gamma_\nu\}=2\delta_{\mu\nu}$.
The gauge coupling constant is defined by $g^2=2e^2/\epsilon_0(1+\varepsilon)$,
with the electric charge $e$, the dielectric constant of vacuum $\epsilon_0$, and the dielectric constant of substrate $\varepsilon$.
 Due to the small Fermi velocity of quasiparticles, $v_{_F}=3.02\times 10^{-3}$,
 one may adopt the ``instantaneous approximation''
 in which  the spatial component $\mathbf{A}$ is neglected.
%XXXXX [discussion on chiral symmetry here ] XXXXX
 From the absence of the $z$-component in the Dirac operator,
this model possesses a continuous global U(4) symmetry
generated by 16 generators $\{1,\gamma_3,\gamma_5,\gamma_3 \gamma_5\} \otimes \{1,\vec{\sigma}\}$,
which is the extension of the continuous U(1) charge symmetry
and the discrete $Z_2$ sublattice exchange symmetry of the original honeycomb lattice.
The explicit chiral symmetry breaking term is represented by
the mass $m$ in Eq.(\ref{eq:effaction}).
 
After  performing  the scale transformation in the temporal direction,
$x_4 \rightarrow x_4 /v_{_F}, \; A_4 \rightarrow v_{_F} A_4$,
 the Dirac operator becomes independent of $v_{_F}$.
 This scale transformation changes the mass $m$ into the effective mass $m_*=m/v_{_F}$,
 while the Coulomb coupling strength is enhanced as $g_*^2 = g^2/v_{_F}$
which is about 300 times larger than that of QED.
Since the inverse effective coupling strength 
$\beta=1/g_*^2$ is $0.0369$ in the vacuum-suspended graphene,
 the expansion by $\beta$ (strong coupling expansion) would work well.

\subsection{Regularization on a square lattice}
 We discretize the low energy action Eq.(\ref{eq:effaction}) on a square
 lattice with a lattice spacing $a$. 
 Since  the original honeycomb lattice has 
 a lattice spacing $a_{_\mathrm{Hc}} \sim 1.4$\AA, 
  we make an approximate identification, $a \sim a_{_\mathrm{Hc}}$, so that
 we can carry out the strong coupling expansion. 
 The quasiparticles in monolayer graphene are
  described with a single staggered fermion $\chi$,
because its eight doublers can be identified as 4(spinor components) $\times$ 2(spin) degrees of freedom.
As a consequence, the lattice action for fermions on graphene is written as \cite{drut_2009}
\begin{equation}
S_F= \sum_{x^{(3)}}  \left[  \frac{1}{2} \sum_{\mu=1,2,4}
 \left( V_{\mu}^+(x)-V_{\mu}^-(x) \right)  + m_{*} M(x) \right]
\label{eq:latticeaction-F}
\end{equation} 
with fermionic bilinears
$M(x)= \bar{\chi}(x)\chi(x), V_{\mu}^+(x)=  \eta_{\mu}(x)\bar{\chi}(x)U_{\mu}(x) \chi(x+\hat{\mu}), V_{\mu}^-(x)=\eta_{\mu}(x)\bar{\chi}(x+\hat{\mu}) U_{\mu}^{\dagger}(x) \chi(x)$,
where $\mu=1,2,4$.
The staggered phase factors  $\eta_{\mu}$ 
  corresponding to the Dirac $\gamma$-matrices are
$ \eta_4(x)=1, \eta_1(x)=(-1)^{x_4},  \eta_2(x)=(-1)^{x_4+x_1}$.
$U_\mu$ is the U(1) link variable,
where the temporal link is defined as $U_4(x)=\exp\left[i\theta(x)\right] \, (-\pi \leq \theta < \pi)$, while  the spatial links $U_{1,2,3}$ are set to unity as a result of the instantaneous approximation.  In the staggered fermion formulation,
 the global chiral symmetry U(4) shrinks to $\mathrm{U(1)_V \times U(1)_A}$,
with the ordinary $\mathrm{U(1)_V}$ charge symmetry,
and the axial $\mathrm{U(1)_A}$ symmetry generated by $\epsilon(x) \equiv (-1)^{x_1+x_2+x_4}$.

As for the pure gauge action,
we consider two types of formulation.
One is the compact formulation
which consists of plaquettes made of U(1) compact link variables:
\begin{equation}
S_G^{\rm (C)} = \beta \sum_{x^{(4)}} \sum_{j=1,2,3}
\left[1-{\rm Re} \left( U_4(x) U_4^{\dagger}(x+\hat{j}) \right)\right].
 \label{eq:latticeaction-G}  
\end{equation}
The other is the non-compact formulation
with  the gauge angle  $\theta$:
\begin{equation}
S_G^{\rm (NC)} = \frac{\beta}{2} \sum_{x^{(4)}} \sum_{j=1,2,3} \left[\theta(x)-\theta(x+\hat{j})\right]^2. \label{eq:latticeaction-G-NC}
\end{equation}
The compact formulation has photon self-interactions 
which are absent in the non-compact formulation and in the continuum theory. 

\section{Strong coupling expansion}
\begin{figure}[htb]
\vspace{0.5cm}
\begin{center}
\includegraphics[width=7cm]{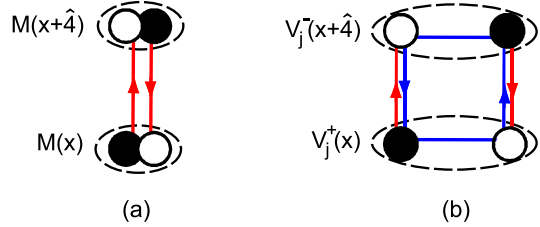}
\end{center}
\vspace{-0.5cm}
\caption{Induced four-fermion interaction in the strong coupling
 expansion.
The open and filled circles represent
 fermion fields $\chi$ and $\bar{\chi}$, respectively.
(a) In the LO,  the time-like links (red arrows) 
in the fermion action $S_F$ 
 cancel with each other to leave a spatially local interaction.
(b) In the NLO,
 the time-link links in $S_F$ are canceled by the time-like links in 
 a plaquette $S_G$  (blue arrows)  to leave a spatially non-local interaction.
}
\label{fig:links}
\end{figure}

Expanding the partition function $Z$ by the inverse coupling strength $\beta$
and integrating out the link variables order by order,
we obtain the effective action $S_\chi$  in terms of fermions \cite{Drouffe:1983fv}:
\begin{equation}
Z= \int [d\chi d\bar{\chi}][d\theta] \left[ \sum_{n=0}^{\infty}
 \frac{(-S_G)^n }{n!} e^{-S_F} \right] = \int
 [d\chi d\bar{\chi}] e^{-S_{\chi}}.
 \label{eq:part-Z}
\end{equation}
Since the link integration selects the terms
in which the link variable cancel with its complex conjugate, various
4-fermi interaction terms are induced as shown in Fig.\ref{fig:links}.
With the compact formulation,
we obtain the leading order (LO) and the next-to-LO (NLO) effective action,
$S_\chi^{(0)}$ and $S_{\chi}^{(1) {\rm C}}$ respectively, as
\begin{eqnarray}
S_{\chi}^{(0)} &=& \sum_{x^{(3)}}  \left[ \frac{1}{2} \sum_{j=1,2} 
  \left( V_j^+(x)-V_j^-(x) \right)  + m_{*} M(x) \right] -\frac{1}{4} \sum_{x^{(3)}} M(x) M(x+\hat{4}), \label{eq:4-fermi-0} \\
S_{\chi}^{(1) {\rm C}} &=& \frac{\beta}{8} \sum_{x^{(3)}} \sum_{j=1,2}
\left[ V_j^+(x) V_j^-(x+\hat{4}) + V_j^-(x) V_j^+(x+\hat{4}) \right]. \label{eq:4-fermi-1}
\end{eqnarray}
Since the pure gauge term $S_G$ vanishes in the strong coupling limit ($\beta=0$),
the compact formulation and the non-compact one give the same result in the LO.
In the NLO,
the effective action from the non-compact formulation, $S_{\chi}^{(1) {\rm NC}}$,
is twice that from the compact one, $S_{\chi}^{(1) {\rm C}}$, so that
 observables of the both formulations are related as
 ${\cal O}^{{\rm NC}}(\beta)={\cal O}^{{\rm C}}(2\beta)$. 

\begin{figure}[htb]
\begin{center}
\includegraphics[width=8cm]{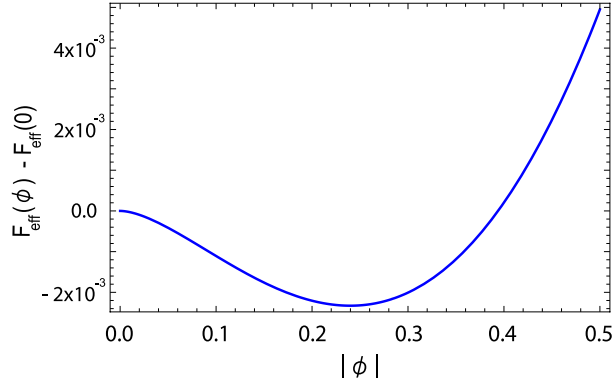}
\end{center}
\vspace{-0.5cm}
\caption{The free energy in the strong coupling limit, $F_{\rm eff}^{(0)}(\phi)$,
in the lattice unit as a function of $|\phi|$, in the chiral limit ($m=0$).}
\label{fig:NLO}
\end{figure}

In order to linearize the induced 4-fermi terms and to integrate out the fermions,
we introduce complex bosonic auxiliary fields by the Stratonovich--Hubbard transformation.
As for the LO 4-fermi terms in Eq.(\ref{eq:4-fermi-0}),
we introduce an auxiliary field $\phi(x) = \phi_\sigma(x)+i\epsilon(x)\phi_\pi(x)$.
Another auxiliary field $\lambda=\lambda_1+i\lambda_2$ is introduced
 to linearize  the NLO terms in Eq.(\ref{eq:4-fermi-1}); 
 the mean field value of $\lambda$
  is determined by requiring the stationary condition of the effective action.
% By applying mean-field approximation over $\phi_\sigma$ and $\phi_\pi$,
Then we arrive at the LO and the NLO effective potential (free energy) 
written in terms of $\phi$ as
\begin{eqnarray}
F_{\rm eff}^{(0)}(\phi) &=& \frac{1}{4}|\phi|^2 -\frac{1}{2}\int_\mathbf{k} \ln\left[G^{-1}(\mathbf{k};\phi)\right],
\label{eq:effpot-LO} \\
F_{\rm eff}^{(1){\rm C}}(\phi) &=& -\frac{\beta}{4} \sum_{j=1,2} \left[\int_\mathbf{k} G(\mathbf{k};\phi) \sin^2 k_j\right]^2. \label{eq:effpot-NLO}
\end{eqnarray}
with the bosonic effective propagator defined as
$G^{-1}(\mathbf{k};\phi) \equiv \left|\phi/2-m_*\right|^2 +\sum_{j=1,2}\sin^2 k_j$
and the momentum integration as $\int_\mathbf{k} \equiv (2\pi)^{-2} \int_{-\pi}^{\pi} d k_1 \int_{-\pi}^{\pi} d k_2$.
Fig.\ref{fig:NLO} shows $F_{\rm eff}^{(0)}$ in the chiral limit ($m=0$):
  Its minimum corresponds to the chiral condensate
(the order parameter of chiral symmetry breaking)
$\sigma \equiv |\langle \bar{\chi}\chi \rangle| = |\langle\phi\rangle|$,
 so that the spontaneous ``chiral symmetry'' breaking takes place
in the strong coupling limit.
The effective mass of fermions  reads $M_F = m+(v_{_F}/a)(\sigma a^2/2)$,
from the mass term of the effective action.

Since $F_{\rm eff}^{(1)}$ is an increasing function of $|\phi|$,
the chiral condensate $\sigma$ drops as $\beta$ grows.
In other words, the chiral symmetry tends to be
 restored as the coupling strength becomes weaker.
Up to the linear terms in $\beta$ and $m$, we can calculate $\sigma$ with the compact formulation as
\begin{equation}
\sigma^{{\rm C}}(\beta,m) \simeq (0.240 - 0.297 \beta + 0.0239\ ma) a^{-2}, \label{eq:sigma-exp}
\end{equation}
As we mentioned,  the condensate in
 the non-compact formulation is simply obtained as
  $\sigma^{{\rm NC}}(\beta) = \sigma^{{\rm C}}(2\beta)$ up to NLO.
The behavior that $\sigma^{{\rm NC}}(\beta)$ drops faster than $\sigma^{{\rm C}}(\beta)$
is consistent with the results of the 
 Monte Carlo simulation for the same lattice model \cite{drut_2010}.
Taking $a^{-1}\simeq a^{-1}_{_{\rm Hc}} = 1.39 \ {\rm keV}$ as a typical cutoff scale of our system,
we obtain $\sigma^{{\rm C}}(\beta,m) \simeq \left[\left(0.680 - 0.421 \beta + \frac{1.39\ m}{\rm eV} \right) {\rm keV} \right]^2$.
The dynamical fermion mass is estimated from the value of the chiral condensate as
$M_F \simeq (0.523 - 0.623 \beta ) \ {\rm eV} \ + 3.05 m$,
which is much smaller than the momentum cutoff scale $E_\Lambda \sim \pi v_{_F}/a = 13\mathrm{eV}$ of the original honeycomb lattice 
as long as the bare mass $m$ is small enough.

\section{Collective Excitations}
Here we study the fluctuations of the order parameter $\phi(x)$
around the symmetry broken state $\langle \phi \rangle = -\sigma$:
the phase fluctuation (``$\pi$-exciton'') corresponds to $\phi_\pi(x)$
and the amplitude fluctuation (``$\sigma$-exciton'')
 corresponds to $\phi_\sigma(x)$.
Propagators of those modes, $D_{\phi_{\sigma,\pi}}$ are derived
 from the second derivative
of the effective action $S_{\rm eff}[\phi]$ with 
respect to the corresponding fields $\phi_{\sigma,\pi}$.
 Their excitation energies are obtained from the imaginary pole of the propagator,
${D}_{\phi_{\sigma,\pi}}^{-1}(\mathbf{p}=0,\omega=i M_{\sigma,\pi}/v_{_F})=0$.

As for the $\pi$-exciton, we obtain a mass formula in the leading order of $m$ as
\begin{eqnarray}
 M_\pi \simeq \frac{2 v_{_F}}{a} \sqrt{\frac{m}{M_F(m=0)}}. \label{eq:mpi}
\end{eqnarray}
Since $M_\pi$ vanishes in the chiral limit ($m=0$),
 this mode serves as a pseudo-Nambu--Goldstone (NG) boson
emerging from the spontaneous breaking of chiral  symmetry.
 From the axial Ward--Takahashi identity
corresponding to the infinitesimal local chiral transformation,
a simple relation  can be derived,
$(F^{\tau}_{\pi}{M_\pi})^2 = m \sigma$,
which is analogous to the Gell-Mann--Oakes--Renner relation for the pion in QCD \cite{gmor}.
Here the temporal ``pion decay constant'' $F_{\pi}^{\tau}$ is defined
 by the matrix element, $\langle 0 | J^{\rm axial}_4(0)| \pi(p) \rangle = 2  F_{\pi}^{\tau} p_{\pi}^\tau$,
with the axial current $J^{\rm axial}_{\mu}(x) \equiv \frac{i}{2} \epsilon(x) \left[V_{\mu}^-(x) - V_{\mu}^+(x)\right]$.

By solving the pole equation numerically, the $\sigma$-exciton is found to
be  a massive mode with $M_{\sigma} \simeq (1.30 - 0.47 \beta)({v_{_F}}/{a})+22.6m$.
Since $M_{\sigma}$ acquires a large value comparable to the cutoff scale $E_\Lambda$,
 application of the low-energy effective theory in this channel is not 
 quite justified.

\section{Conclusion}
We have investigated the behavior of monolayer graphene  analytically 
in/around the limit of strong Coulomb coupling,
by means of the strong coupling expansion of U(1) lattice gauge theory.
As for the pure gauge action, we have compared the results from the 
compact and  non-compact formulations. In either case,
 we find that  ``chiral symmetry'' (the sublattice exchange symmetry in the 
  original  honeycomb lattice) is spontaneously broken in the strong coupling limit
with equal magnitude of the chiral condensate.
As the coupling strength becomes weaker, chiral condensate from
 the non-compact formulation drops faster
than that from the compact one.
These results up to NLO  in the strong coupling expansion agree qualitatively with
the extrapolation of the numerical results of the 
lattice Monte Carlo simulations.

We have also examined the 
collective excitations associated with the chiral symmetry breaking in our approach
and have derived their dispersion relations.
The phase fluctuation of the chiral condensate, the ``$\pi$-exciton'',
 behaves as a pseudo-NG boson, like the pion in QCD.
 Experimental observation of such mode in vacuum-suspended graphene
would be a good evidence for the spontaneous chiral symmetry breaking
 in the strong coupling regime.
The amplitude fluctuation of the chiral condensate, the  ``$\sigma$-exciton,''
acquires  a large mass comparable to the intrinsic cutoff scale $E_\Lambda$, 
 so that it  needs further investigation without the low-energy approximation.

 There are several directions to be investigated in future:
Behavior of the present model on a square lattice  
 at  finite temperature and  finite chemical potential
 still remains an open problem  both analytically and numerically.
 Formulating the strong coupling expansion  on a 
  honeycomb lattice  would be of great importance.
  Extension of our strong coupling approach to the analysis of
   bilayer graphene, 
which has been attracting attentions recently \cite{Guinea_2010},
    would be also of interest. 

\section*{Acknowledgements}
The authors thank H. Aoki, T.Z. Nakano, Y. Nishida, 
 A. Ohnishi, T. Oka,  S. Sasaki, E. Shintani and  N. Yamamoto
  for discussions.
Y.~A.~ is supported by Grant-in-Aid for Japan Society for the Promotion of Science (DC1, No.22.8037). T. H. is
supported in part by the Grant-in-Aid for Scientific Research on Innovative Areas
(No. 2004: 20105003) and by Japanese MEXT grant (No. 22340052).

\end{document}